\documentclass[conference]{IEEEtran}

\IEEEoverridecommandlockouts

\usepackage{caption}
\usepackage{subcaption}
\usepackage{graphicx}
\usepackage{cite}
\usepackage{url}
\usepackage[cmex10]{amsmath}
\usepackage{amssymb}
\usepackage{algorithm}
\usepackage{algpseudocode}
\usepackage{cases}
\usepackage{color}
\usepackage{soul}
\usepackage{xcolor}
\usepackage{framed}
\usepackage{cite}
\usepackage{epstopdf}
\usepackage{calc}
\usepackage{array}
\usepackage{amsmath}
\usepackage{comment}
\DeclareGraphicsExtensions{.pdf,.eps,.png,.jpg,.mps}

\begin{document}
\title{Feasibility of Serving $K$ Secondary Users in Underlay Cognitive Radio Networks using Massive MIMO}
\author{\IEEEauthorblockN{Shailesh Chaudhari and Danijela Cabric}
\IEEEauthorblockA{Department of Electrical Engineering, University of California, Los Angeles\\
Email: schaudhari@ucla.edu, danijela@ee.ucla.edu}\\

\thanks{This work has been supported by the National Science Foundation under grant 1149981.}
\vspace{-10mm}}
\maketitle


\begin{abstract}
In this paper, we analyze the feasibility of serving $K$ secondary users (SUs) on the downlink using a secondary base station (SBS), equipped with a large antenna array, in an underlay cognitive radio (CR) network. First, we formulate a feasibility problem in order to compute beamforming vectors and power allocations for $K$ SUs with constraints on the maximum
 allowable interference to primary users (PUs), required minimum rate at SUs, and maximum transmit power from SBS. 
 The problem formulation takes into account the imperfect channel state information between SUs and PUs. We propose a two step approach to solve the non-convex problem. In the first step, beamforming vectors are computed using one of the two alternative schemes: maximum eigenmode beamforming (MEB) or zero forcing beamforming (ZFB). We show that the power allocations can be computed by solving a linear programming feasibility problem. Further, we provide theoretical feasibility analysis of serving $K$ SUs with equal power allocation by deriving the distributions of SINR at SUs and interference to PUs, thereby computing the probability of serving $K$ SUs as a function of the constraints. The analytical and simulation results are presented to demonstrate the impact of constraints on  the feasibility of serving $K$ SUs in the CR network.\\
\end{abstract}
\IEEEpeerreviewmaketitle


\section{Introduction}
\label{sec:Introduction}

Due to increasing number of wireless devices and limited spectrum availability, there is a need to efficiently use the available radio spectrum and serve multiple devices (users) at the same time. The cognitive radio (CR) network utilizes the spectrum more efficiently by allowing spectrum sharing among licensed primary users (PUs) and unlicensed secondary users (SUs). Especially in underlay CR networks, SUs are allowed to transmit concurrently with PUs in the same spectrum, as long as the interference to PU receiver is kept below a specified limit \cite{Goldsmith2012}. The SUs in the underlay CR network employ MIMO beamforming techniques in order to limit the interference at PUs, thereby effectively limiting the probability of outage at PUs. Further, multiple SUs can be, simultaneously, served in the CR network by utilizing a massive MIMO system, where the secondary transmitter is equipped with a large antenna array \cite{Asilomar2015}. 

Various recent works in the literature have studied the underlay MIMO CR system \cite{Yiu2012, Wang2015, Zhang2015, Du2012,Du2013}. 
In order to maximize the achievable rate at SUs and  contain the interference to PUs, works in \cite{Yiu2012} and \cite{Zhang2015} assume perfect channel state information (CSI) between SUs and PUs. However, such a perfect knowledge is not always available in the CR network, especially if PUs are non-cooperative.  Further, the work in \cite{Zhang2015} does not take into account the rate constraints on individual SUs. Numerical results in \cite{Du2012} show that the beamforming and power allocation become infeasible with high probability as the rate constraints at the SUs increases. However, there is no comprehensive analysis available in the literature that addresses the question of feasibility of serving SUs with certain minimum rate, while keeping the interference to PUs  below a specified limit using only the imperfect CSI between SUs and PUs.

In order to address this question, we first formulate a feasibility problem to compute beamforming vectors and power allocations in order to serve $K$ SUs on the downlink using a secondary base station (SBS), equipped with a large antenna array. The constraints of the problem include the maximum allowable interference to PUs, required minimum rate at SUs, and maximum transmit power from the SBS.  We propose a two step approach, comprising of beamforming and power allocation steps, to solve the feasibility problem. Two alternative beamforming techniques are considered, namely maximum eigenmode beamforming (MEB) which requires no CSI between SUs and PUs, and zero forcing beamforming (ZFB) which works with imperfect CSI. 
We analyze the feasibility of MEB and ZFB under equal power allocation and compute the  probability of serving $K$ SUs as a function of the constraints. The analysis presented in this work is helpful in selecting the beamforming technique, based on the availability of CSI between SUs and PUs, as well as selecting the number of antennas at SBS in order to serve $K$ SUs with given constraints.

The paper is organized as follows. The system model and problem formulation are presented in Section \ref{sec:Model}. Section \ref{sec:Beamforming} describes the beamforming techniques and power allocation.  The probability of serving $K$ SUs is derived in Section \ref{sec:Distributions}. Simulation results are presented in Section \ref{sec:Results}, while conclusions and future work are discussed in Section \ref{sec:Conclusion}.

Notations: Lowercase and uppercase bold letters indicate vectors and matrices, respectively, e.g., $\mathbf{\mathbf{v_k}, H_{k}}$. Scalar variables are denoted by non-bold letters, e.g., $\epsilon,P_k$. $\Gamma(k,\theta)$ indicates Gamma distribution with shape parameter $k$ and scale parameter $\theta$, while $\Gamma(k)$ indicates Gamma function and $\gamma(k,x)$ is lower incomplete Gamma function. 

\section{System Model and Problem Formulation}
\label{sec:Model}
\begin{figure}
	\centering
	\includegraphics[width= \columnwidth]{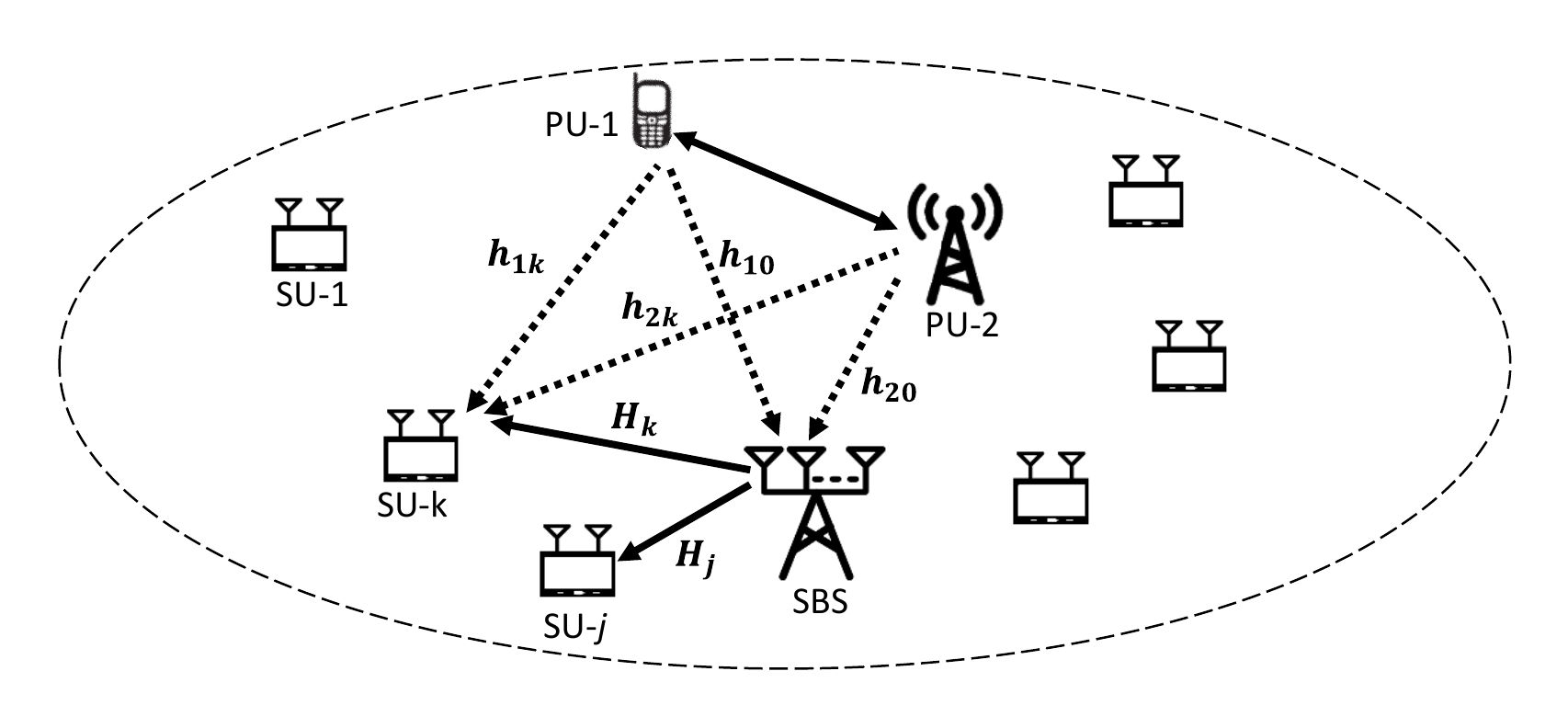}
	\caption{System model: Underlay cognitive radio network.}
	\label{fig:schematic}
	\vspace{-5mm}
\end{figure}
The underlay CR network, as shown in Fig. \ref{fig:schematic}, consists of a SBS with $M_b$ antennas, $K$ SUs each with $M_u$ antennas ($K, M_u\ll M_b$), and $L$ PUs. Since the number of antennas at the SBS is order of magnitude larger than the number of SUs, the network is considered as massive MIMO CR network. The PUs are assumed to be single antenna terminals. Further, let $\mathcal{L}_{tx}$ and $\mathcal{L}_{rx}$ be the set of indices indicating PU transmitters and PU receivers, respectively. Let $|\mathcal{L}_{tx}| = L_{tx}, |\mathcal{L}_{rx}| = L_{rx}$ be the cardinalities of the sets such that ${L}_{tx}+{L}_{rx}=L$. The MIMO channel between SBS and SU-$k$ is denoted by $\mathbf{H_k}$. The SIMO channel between PU-$l$ and SU-$k$ is denoted by $\mathbf{h_{lk}}$, while the channel between PU-$l$ and SBS is $\mathbf{h_{l0}}$. We consider Rayleigh fading propagation model. The elements of channel matrices are iid, complex Gaussian random variables with zero mean and variance $\sigma^2_h$: $CN(0, \sigma^2_h)$ \cite{Yang2013}. The dimensions of matrices $\mathbf{H_k, h_{lk}}$, and  $\mathbf{h_{l0}}$ are $M_u \times M_b, M_u \times 1$, and $M_b \times 1$, respectively. The estimated channels to PUs at SBS and SU are denoted by $\mathbf{\hat{h}_{l0}}$ and $\mathbf{\hat{h}_{lk}}$, respectively. The estimation error $\mathbf{\delta_{l0}} = \mathbf{h_{l0}} - \mathbf{\hat{h}_{l0}}$ is modeled as a complex Gaussian random variable with zero mean and variance $\sigma^2_\delta$. i.e. $\mathbf{\delta_{l0}} \sim CN(0,\sigma^2_\delta \mathbf{I_{M_b}})$, where $\mathbf{I_{M_b}}$ is an identity matrix of size $M_b \times M_b$. Similarly, $\mathbf{\delta_{lk}} =  \mathbf{h_{lk}}-\mathbf{\hat{h}_{lk}}  \sim CN(0,\sigma^2_\delta \mathbf{I_{M_u}})$. The channels between any two nodes in the network are assumed to be reciprocal. 

Let $\mathbf{v_k}$ and $\mathbf{u_k}$ be the unit power transmit and receive beamforming vectors corresponding to the link between SBS and SU-$k$ and $P_k$ be the power allocated to the downlink stream transmitted to SU-$k$. Therefore, the interference to PU-$l$ due to downlink transmission from SBS to $K$ SUs is:
{\small
\begin{equation}
I^{l}_{SU-PU} = \sum_{k=1}^{K} P_k |\mathbf{v^H_k} \mathbf{h_{l0}}|^2,~ \forall l \in \mathcal{L}_{rx} 
\label{eq:int_to_pu}
\end{equation}
}
Similarly, the interference received at SU-$k$ from PU transmitter is 
$
I^{k}_{PU-SU} = \sum_{l\in \mathcal{L}_{tx}} P_p |\mathbf{u^H_k} \mathbf{h_{lk}}|^2,
$
where $P_p$ is the power transmitted by the PU. The interference power at SU-$k$ due to the stream transmitted to SU-$j$, $j\neq k$ (inter-stream interference) is 
$
I^{k}_{SU-SU} = \sum_{j=1, j \neq k}^{K} P_j |\mathbf{u^H_k} \mathbf{H_k} \mathbf{v_j}|^2
$.
Therefore, the SINR at SU-$k$ is expressed as:
{\small
\begin{align}
	SINR_k = \frac{P_k |\mathbf{u^H_k} \mathbf{H_k} \mathbf{v_k}|^2}{\sigma^2_w + I^{k}_{PU-SU} + I^{k}_{SU-SU}},
	\label{eq:sinr_k}
\end{align}
}
where $\sigma^2_w$ is the noise power at SU-$k$. 


It should be noted that the SBS has imperfect knowledge of the channels to PUs. Therefore, the estimated interference to PUs $\hat{I}^{l}_{SU-PU}$ is different from the true interference, ${I}^{l}_{SU-PU}$. In this paper, the estimated interference is modeled as the expected value of the interference given the estimated channels $\mathbf{\hat{h}_{l0}} = \mathbf{h_{l0}} + \mathbf{\delta_{l0}}$:
{\small
\begin{align}
\small \hat{I}^{l}_{SU-PU} &= E\left[P_k |\mathbf{v^H_k} \mathbf{h_{l0}}|^2 |\mathbf{ \hat{h}_{l0}}\right]
= \sum_{k=1}^{K} P_k \left(|\mathbf{v^H_k} \mathbf{\hat{h}_{l0}}|^2 +\sigma^2_\delta\right).
\label{eq:est_int_to_pu}
\end{align}
}
Similarly, the estimated interference power from PU transmitter at SU-$k$ is expressed as
{\small	$\hat{I}^{k}_{PU-SU}  = \sum_{l \in \mathcal{L}_{tx}} P_p \left(|\mathbf{u^H_k} \mathbf{\mathbf{\hat{h}_{lk}}}|^2 +\sigma^2_\delta\right),$}
and the estimated SINR at SU-$k$ is then
{\small
\begin{align}
\small	\widehat{SINR}_k = \frac{P_k |\mathbf{u^H_k} \mathbf{H_k} \mathbf{v_k}|^2}{\sigma^2_w + \sum \limits_{l \in \mathcal{L}_{tx}} P_p \left(|\mathbf{u^H_k} \mathbf{\hat{h}_{lk}}|^2 +\sigma^2_\delta\right) + I^{k}_{SU-SU}}.
	\label{eq:est_sinr_k}
\end{align}
}
Therefore, the feasibility problem should take into account the estimated SINR and interference to PUs. The feasibility problem is expressed as:
{\small
\begin{align}
\text{Find}~~ \{\mathbf{v_k}, \mathbf{u_k}, P_k\}, & \forall {k=1,2,...,K}
\label{eq:fp_start2}
\\\text{Subject to}:
\hat{I}^{l}_{SU-PU} &\leq I^0, \forall l \in \mathcal{L}_{rx}
\label{eq:fp_int_const2}
\\\log_2(1+\widehat{SINR}_k) &\geq R^0, \forall k
\label{eq:fp_rate_const2}
\\\sum_{k=1}^{K} P_k &\leq P^0,~P_k \geq 0
\label{eq:fp_pow_const2}
\\ ||\mathbf{u_k}||^2 = 1, & ~||\mathbf{v_k}||^2 = 1, \forall k,
\label{eq:fp_end2}
\end{align}
}
where $I^0$ is the maximum allowable interference to PUs, $R^0$ is the minimum required rate at SUs, and $P^0$ is the maximum transmit power from SBS. The feasibility problem  (\ref{eq:fp_start2})-(\ref{eq:fp_end2}) is a non-convex problem. We observe that the power constraints (\ref{eq:fp_pow_const2}) and interference constraints (\ref{eq:fp_int_const2}) require $P_k's$ to be small, while the rate constraints (\ref{eq:fp_rate_const2}) require $P_k's$ to be high. Now, let us consider the terms $T_1 = \frac{|\mathbf{u^H_k} \mathbf{H_k} \mathbf{v_k}|^2}{\sigma^2_w + \hat{I}^{k}_{PU-SU} + I^{k}_{SU-SU}}$ in the SINR (\ref{eq:est_sinr_k}) and $T_2 = |\mathbf{v^H_k} \mathbf{\hat{h}_{l0}}|^2$ in the interference (\ref{eq:est_int_to_pu}). Note that the size of feasible set for $P_k$ becomes larger as $T_1$ increases and $T_2$ decreases. Therefore, a feasible power allocation can be computed by designing the beamforming vectors $\mathbf{u_k, v_k}$ so as to obtain a large value of $T_1$ and a small value of $T_2$. With this in mind, in order to solve the non-convex feasibility problem, we divide the problem in two steps: 1) beamforming to compute $\mathbf{v_k},\mathbf{u_k}$ while maximizing $T_1$ and minimizing $T_2$, 2) power allocation to compute $P_k$. Two beamforming techniques are considered, namely, maximum eigenmode beamforming (MEB) and zero forcing beamforming (ZFB). In MEB, $T_1$ is maximized 
 by aligning $\mathbf{u_k}$ and $\mathbf{v_k}$ to principal left and right singular vectors of the channel matrix $\mathbf{H_k}$, while in ZFB, $T_2$ and $I^k_{SU-SU}$ are minimized (i.e. set to zero). Once $\mathbf{u_k, v_k}$ are computed using either MEB or ZFB, the power allocation problem is reduced to a linear feasibility problem. In order to compute power allocations, we investigate two alternative approaches. In the first approach, the power allocations are obtained by directly solving the linear feasibility problem, while in the second approach equal power is allocated to all SUs. The equal power allocation scheme is of interest especially when large scale fading coefficients are assumed to be equal for all the channels in the network such as in this paper and in \cite{Du2012, Yang2013, Du2013}. The beamforming techniques and power allocation schemes are explained in the next section.

\section{Beamforming and Power Allocation}
\label{sec:Beamforming}

\subsection{Maximum eigenmode beamforming (MEB)}
\label{sec:MEB}
In MEB, the $\mathbf{u_k}$ and $\mathbf{v_k}$ are the principal left and right singular vectors of the channel matrix $\mathbf{H_k}$. If singular value decomposition (SVD) of $\mathbf{H_k} = \sum_{i=1}^{M_u}\sigma_{k,i}\mathbf{u_{k,i}v^H_{k,i}}$, where the singular values are: $\sigma_{k,1} \geq \sigma_{k,2} \geq ...\geq \sigma_{k,M_u}$. Then we have: 
$
\mathbf{u_k}=\mathbf{u_{k,1}}~~\text{and}~~ \mathbf{v_k} = \mathbf{v_{k,1}}.
$
 It should be noted that in MEB, the SBS and SUs do not use the knowledge of channels to/from PUs in order to compute $\mathbf{u_k}$ and $\mathbf{v_k}$. Further, the channel gain after beamforming is {\small$|\mathbf{u^H_{k,1}}\mathbf{H_k} \mathbf{v_{k,1}}|^2=\sigma^2_{k,1}$}. Note that the value of $\sigma^2_{k,1}$ increases with increased number of antennas. Therefore, the massive MIMO link between SBS and SU-$k$ achieves large beamforming gain through the term {\small $|\mathbf{u^H_{k,1}}\mathbf{H_k} \mathbf{v_{k,1}}|^2$.}
 
 Further, the interference received from PU-$l$ at SU-$k$ becomes a constant for a given channel realization once the beamforming vector $\mathbf{u_{k,1}}$ is computed {\small $\hat{I}^{k}_{PU-SU}= \sum_{l \in \mathcal{L}_{tx}} P_p (|\mathbf{u^H_{k,1}} \mathbf{\mathbf{\hat{h}_{lk}}}|^2 +\sigma^2_\delta)$}. The inter-stream interference is {\small $I^{k}_{SU-SU} = \sum_{j=1, j \neq k}^{K} P_j |\mathbf{u^H_{k,1}} \mathbf{H_k} \mathbf{v_{j,1}}|^2 = \sigma^2_{k,1} \sum_{j=1, j \neq k}^{K} P_j |\mathbf{v^H_{k,1}} \mathbf{v_{j,1}}|^2$}. Therefore, the feasibility problem (\ref{eq:fp_start2}) - (\ref{eq:fp_end2}) reduces to the following linear feasibility (LF) problem with variables $P_k$. The problem is denoted by LF MEB: 
{ \small
 \begin{align}
 & \textbf{LF MEB:} \text{~~Find}~~ \{P_k\},  \forall {k=1,2,...,K,}
 \label{eq:lp_start}
\\& \nonumber \text{s.t.}:
 \hat{I}^{l}_{SU-PU} = \sum_{k=1}^{K}P_k (|\mathbf{v^H_{k,1} \hat{h}_{l0}}|^2 + \sigma^2_\delta)  \leq I^0, \forall l \in\mathcal{L}_{rx}
\\&\nonumber \frac{\sigma^2_{k,1}P_k}{2^{R_0}-1} \geq \left[ \sigma^2_w +\hat{I}^{k}_{PU-SU} + \sigma^2_{k,1} \sum_{j=1, j \neq k}^{K} P_j |\mathbf{v^H_{k,1}} \mathbf{v_{j,1}}|^2 \right], 
\\ &\sum_{k=1}^{K}P_k \leq P^0, P_k \geq 0. 
 \label{eq:lp_end}
 \end{align}
}
The feasible power allocation is obtained using standard linear program solvers. 
\subsection{Zero forcing beamforming (ZFB)}
\label{sec:ZF}
In zero forcing beamforming considered in this paper, the receive beamforming vectors $\mathbf{u_k}$ are same as that of MEB: $\mathbf{u_k} = \mathbf{u_{k,1}}$. On the other hand, the transmit beamforming vectors $\mathbf{v_k}$ are obtained so as to nullify $T_2 = |\mathbf{v^H_k} \mathbf{\hat{h}_{l0}}|^2$ and $I^{k}_{SU-SU}$. In order to nullify $I^{k}_{SU-SU}$, let us consider the equivalent channel between SBS and SU-$k$ after applying the receive beamforming vector $\mathbf{u_k}=\mathbf{u_{k,1}}$. The equivalent channel between SU-$k$ and SU-$k$ is $\mathbf{g_k} = \mathbf{H_k^H} \mathbf{u_{k,1}}$. Therefore, the zero forcing beamformer $\mathbf{v_k}$ can be expressed as: 
{\small
\begin{align}
\mathbf{v_k}  = \mathbf{[G (G^H G)^{-1}]}_k
\label{eq:zf_vk}
\end{align}
}
where, $\mathbf{G=[g_1,g_2,...,g_K, \hat{h}_{10},...,\hat{h}_{\mathcal{L}_{rx}0}]}$ and $[.]_k$ indicates $k^{th}$ column of the matrix. The beamforming vectors in (\ref{eq:zf_vk}) ensure that $T_2 = |\mathbf{v^H_k} \mathbf{\hat{h}_{l0}}|^2=0$ and $I^{k}_{SU-SU}=0$, while the equivalent channel gain between SBS and SU becomes {\small $|\mathbf{u_{k,1}^H} \mathbf{H_k} \mathbf{v_k}|^2=\sigma^2_{k,1}|\mathbf{v^H_{k,1}}\mathbf{v_k}|^2$}. Note that $M_b\times 1$ vector $\mathbf{v_k}$ needs $K-1+L_{rx}$ nulls in order to have $T_2=0$ and $I^{k}_{SU-SU}=0$. Therefore, the condition $M_b > K-1+L_{rx}$ must hold. This condition is generally valid in massive MIMO network. Finally, the power allocation problem for ZFB algorithm reduces to the following linear feasibility problem, denoted by LF ZFB:
{\small
\begin{align}
& \textbf{LF ZFB:} \text{~~Find}~~ \{P_k\},  \forall {k=1,2,...,K}
\label{eq:lp_zf_start}
\\\nonumber \text{s.t.~~} & P_k \geq \frac{(2^{R_0}-1)\left(\sigma_w^2 + \sum \limits_{l \in \mathcal{L}_{tx}} P_p \left(|\mathbf{u^H_k} \mathbf{\hat{h}_{lk}}|^2 +\sigma^2_\delta\right)\right)}{\sigma^2_{k,1}|\mathbf{v_{k,1}^H}\mathbf{v_k}|^2}, \forall k
\\ & \sum_{k=1}^{K}P_k\sigma^2_\delta \leq I^0,~~\sum_{k=1}^{K}P_k \leq P^0, P_k \geq 0.
\label{eq:lp_zf_end}
\end{align}
}
Power allocation $P_k = \frac{(2^{R_0}-1)\left(\sigma_w^2 + \sum \limits_{l \in \mathcal{L}_{tx}} P_p \left(|\mathbf{u^H_k} \mathbf{\hat{h}_{lk}}|^2 +\sigma^2_\delta\right)\right)}{\sigma^2_{k,1}|\mathbf{v_{k,1}^H}\mathbf{v_k}|^2}$ ensures that each SU receives equal rate of $R^0$. We refer to this power allocation as equal rate power allocation. It should be noted that if there exists any feasible power allocation for (\ref{eq:lp_zf_start})-(\ref{eq:lp_zf_end}), then equal rate power allocation is also a feasible power allocation \cite{Asilomar2015}. Therefore, the feasibility problem reduces to checking following simple linear condition:
{\small
\begin{align}
\sum_{k=1}^{K} P_k \leq \min \left(P^0, I^0/\sigma^2_\delta\right).
\end{align}
}
If the above condition is satisfied equal rate power allocation, then it is the required feasible power allocation. 

\section{Probability of serving $K$ SUs}
\label{sec:Distributions}
In this section, we derive expressions for probability of serving $K$ SUs with MEB and ZFB under equal power allocation to all SUs. Let $Q_K$ be the probability of serving $K$ SUs with given rate constraints ($R^0$) and interference constraints ($I^0$). We note that $Q_K$ is the joint probability that the SINR at $K$ SUs is $\geq 2^{R^0}-1$ and interference to PU receivers is $ \leq I^0$. Further, SINR at SUs and interference to PUs are independent of each other since they are functions of independent channels $\mathbf{H_k}$ and $\mathbf{h_{l0}}$. Therefore, the joint probability $Q_K$ can be computed from the marginal distributions of SINR and interference. In order to compute $Q_K$, we first derive the CDFs of SINR at SUs and interference to PUs.

\textbf{Theorem 1:} If equal power $P_k = P^{meb}_{eq}$ is allocated to all SUs, then $SINR^{meb}_k$ is modeled as inverse gamma variable  with shape parameter $k'$ and scale parameter $1/\theta'$ i.e. $SINR^{meb}_k\sim IG(k',1/\theta')$. The parameters and CDF is given below:
{\small
\begin{align}
\nonumber	k' = \frac{(C+A)^2}{B}, ~~ \theta' = \frac{B}{C+A}
\\	Pr(SINR^{meb}_k \leq s) = 1- \frac{\gamma(k',1/s \theta')}{\Gamma(k')} 
\label{eq:inverse_gamma}
\end{align}
} where,
{\small
\begin{align}
\small \nonumber A = \frac{{L}_{tx} P_p \sigma^2_h}{P^{meb}_{eq} E[\sigma^2_{k,1}]} +\frac{K-1}{M_b}, 
B = \frac{{L}_{tx} P^2_p \sigma^4_h}{(P^{meb}_{eq})^2 E[\sigma^2_{k,1}]^2} +\frac{K-1}{M^2_b} 
\end{align}
}
\textit{Proof}: Appendix A.

\textbf{Theorem 2:} Interference to PU in MEB is a Gamma random variable with  shape parameter $K$ and scale parameter $P^{meb}_{eq} \sigma^2_h$: i.e. $I^{l,meb}_{SU-PU} \sim \Gamma(K, P^{meb}_{eq}\sigma^2_h)$ and the CDF is given as
{\small
\begin{align}
Pr(I^{l,meb}_{SU-PU} \leq x) = \frac{\gamma(K, x/P^{meb}_{eq} \sigma^2_h)}{\Gamma(K)}
\label{eq:I_meb_cdf}
\end{align}
}
\textit{Proof}: The interference is expressed as $I^{l,meb}_{SU-PU} = P^{meb}_{eq}\sum_{k=1}^{K}  |\mathbf{v_{k,1}^H} \mathbf{h_{l0}}|^2$. We note that $|\mathbf{v_{k,1}^H} \mathbf{h_{l0}}|^2$ is the power of the projection of a single beamforming vector on an isotropic channel vector $\mathbf{h_{l0}}$. Therefore, $ |\mathbf{v_{k,1}^H} \mathbf{h_{l0}}|^2$ is a Gamma random variable: $ |\mathbf{v_{k,1}^H} \mathbf{h_{l0}}|^2 \sim \Gamma(1,\sigma^2_h)$ \cite{Hosseini2014}. Further, $I^{l,meb}_{SU-PU}$ is scaled sum of $K$ iid Gamma random variables with scaling factor $P^{meb}_{eq}$. Therefore, from lemma of 3 of \cite{Hosseini2014}, we get 
 $I^{l,meb}_{SU-PU} \sim \Gamma(K, P^{meb}_{eq}\sigma^2_h)$.

Now, let us consider $SINR_k$ and $I^{l}_{SU-PU}$ under ZFB with equal power allocation $P_k= P^{zf}_{eq}$. 

\textbf{Theorem 3:} $SINR^{zf}_k$ is modeled as random variable with generalized F distribution: $SINR^{zf}_k \sim GF(k_n, k_d, \lambda)$, where $k_n = M_b -K-L_{tx}+1$ and
{\small
\begin{align}
\nonumber k_d =  \frac{(\sigma^2_w + {L}_{tx}P_p \sigma^2_h)^2}{{L}_{tx} P^2_p\sigma^4_h},~ \lambda = \frac{{L}_{tx} P_p\sigma^2_h M_b}{P^{zf}_{eq}E[\sigma^2_{k,1}](\sigma^2_w + {L}_{tx} P_p \sigma^2_h)} 
\end{align}
}
The CDF of SINR is given as:
{\small
\begin{align}
Pr(SINR^{zf}_{k} \leq s) = \frac{B \left(\frac{\lambda s}{1+ \lambda s},k_n,k_d\right)}{B(k_n, k_d)}
\end{align}
}
where $B(.,.,.)$ and  $B(.,.)$ are incomplete and complete beta functions, respectively \cite{Gia2011}.

\textit{Proof}: Appendix B.

\textbf{Theorem 4:} Interference to PU in ZFB is a Gamma random variable with  shape parameter $K$ and scale parameter $P^{zf}_{eq} \sigma^2_\delta$: i.e. $I^{l,zf}_{SU-PU} \sim \Gamma(K, P^{zf}_{eq}\sigma^2_\delta)$ and the CDF expression is obtained by replacing $\sigma^2_h$ with $\sigma^2_\delta$ in (\ref{eq:I_meb_cdf}).

\textit{Proof}: {\small $I^{l,zf}_{SU-PU} = P^{zf}_{eq}\sum \limits_{k=1}^{K}  |\mathbf{v_{k}^H} \mathbf{h_{l0}}|^2=P^{zf}_{eq}\sum \limits_{k=1}^{K}  |\mathbf{v_{k}^H} \mathbf{\delta_{l0}}|^2 $}. This is due to the fact that $|\mathbf{v^H_{k}}\mathbf{\hat{h}_{l0}}|^2 =0$ and $\mathbf{v^H_{k}}\mathbf{\hat{h}_{l0}}=0$ in zero forcing. Further proof is similar to that of \textit{Theorem 2} since $\mathbf{\mathbf{\delta_{l0}}}$ is an isotropic vector.

Since $SINR^{meb}_k$ and $I^l_{SU-PU}$ are functions of iid channels, we obtain $Q_K$ for MEB by using Theorems 1 and 2 as follows:
{\small
\begin{align}
Q^{meb}_K = 
[Pr(SINR^{meb}_k \geq (2^{R_0}-1))]^K \times
 [Pr(I^{l,meb}_{SU-PU}\leq I^0)]^{{L}_{rx}}
\label{eq:pr_k_served_meb}
\end{align}
}
Similarly, the probability of serving $K$ SUs using ZFB is computed using Theorems 3 and 4 as follows:
{\small
\begin{align}
 Q^{zf}_K =
	[Pr(SINR^{zf}_k \geq (2^{R_0}-1))]^K \times
	 [Pr(I^{l,zf}_{SU-PU}\leq I^0)]^{{L}_{rx}}
\label{eq:pr_k_served_zf}
\end{align}
}
 It should be noted that the power allocations $P^{zf}_{eq}$ and $P^{meb}_{eq}$ must be less than $P^{max}_{eq}=P^0/K$ in order to satisfy the transmit power constraints. Further, in order to obtain a lower bound on equal power allocation, we substitute {\small $I^{k}_{PU-SU} = I^{k}_{SU-SU} =0$, $|\mathbf{u^H_k} \mathbf{H_k} \mathbf{v_k}|^2 = \sigma^2_{k,1}$, and $SINR_k=2^{R^0}-1$} in (\ref{eq:sinr_k}). The minimum equal power allocation is then $P^{min}_{eq} = \sigma^2_w (2^{R^0} -1)/\sigma^2_{k,1}$. The optimum equal power allocation in MEB can be easily found by maximizing (\ref{eq:pr_k_served_meb}) in range $P^{min}_{eq} \leq P^{meb}_{eq}\leq P^{max}_{eq}$. Similarly, the optimum equal power allocation in ZFB is found by maximizing (\ref{eq:pr_k_served_zf}) in range $P^{min}_{eq} \leq P^{zf}_{eq}\leq P^{max}_{eq}$. Note that feasible equal power allocation in ZFB is computed using the channel and estimation error statistics, $\sigma^2_h$ and $\sigma^2_\delta$, while the equal power with MEB uses only channel statistic, $\sigma^2_h$. 
Also note that Theorems 1 and 3 provide expressions for outage probability under MEB and ZEB schemes, respectively.

\section{Simulation Results}
\label{sec:Results}
In the simulation setting, the channel variance is one  ($\sigma^2_h=1$) and noise power is $\sigma^2_w=1$ \cite{Yang2013}. There are $L=2$ PUs in the network with one transmitter with $P_p= 0$ dB{\footnote{ Transmit power and interference are measured in dB relative to the noise power $\sigma^2_w$.}} and one receiver (${L}_{tx}=1$, ${L}_{rx}=1$). The number of antennas at SUs is $M_u=4$. 

\subsection{Impact of power allocation in equal power scheme}
\begin{figure}
	\centering
	\includegraphics[width= \columnwidth]{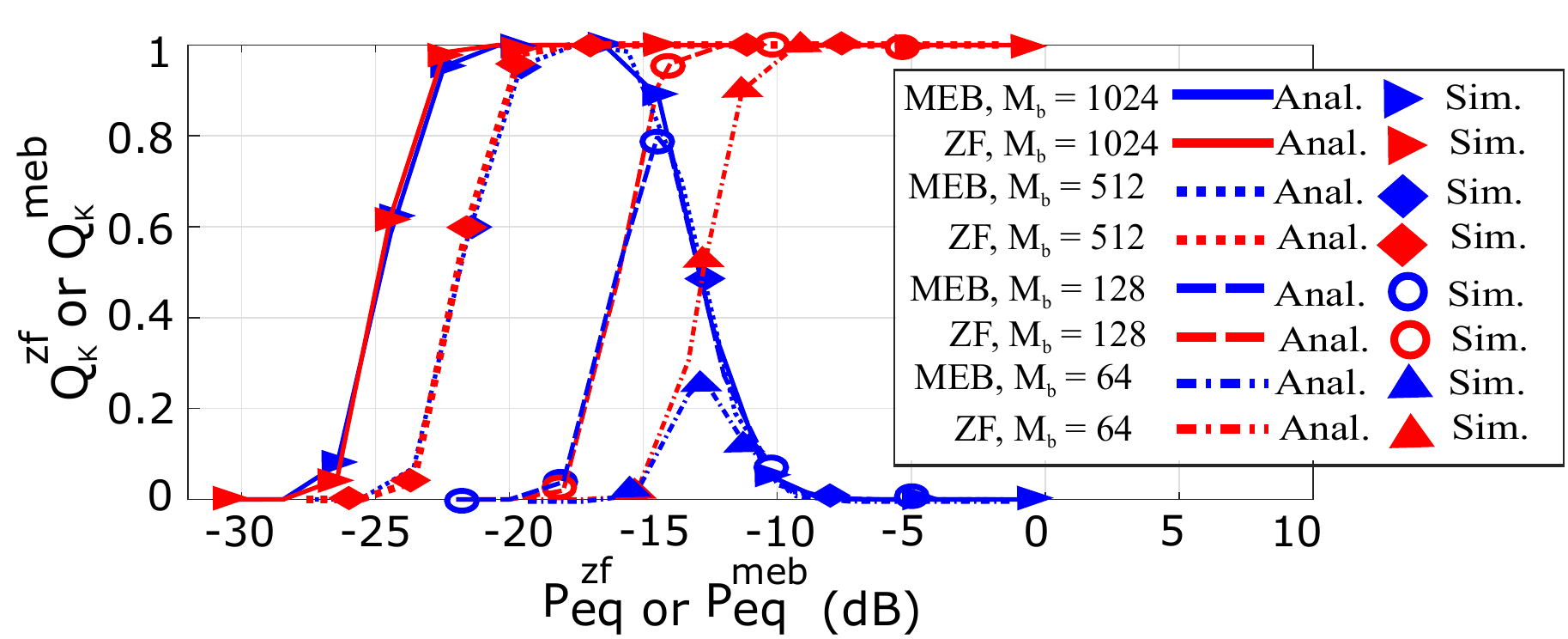}
	\caption{Probability of serving $K=10$ SUs for MEB and ZFB. $I^0=-3$dB, $R^0=1$bps/Hz, $P^0=10$dB, $\sigma^2_\delta=0.01$.}
	\label{fig:eq_pw_Mb}
	\vspace{-5mm}
\end{figure}

First, we study the impact of power allocation in equal power scheme in MEB and ZFB for different number of SBS antennas. The impact of increasing number of antennas on probability of serving $K=10$ SUs is shown in Fig. \ref{fig:eq_pw_Mb}. The maximum equal power with $P^0=10$ dB and $K=10$ is $P^{max}_{eq}=0$ dB. We can see that all 10 SUs are served with probability 1 by ZFB with $M_b = 64$ antennas for $-10$dB $\leq P^{zf}_{eq} \leq 0$dB, while MEB serves 10 SUs with $M_b=64$ with probability 0.27 for $P^{meb}_{eq}=-12.74$ dB. It means that for $M_b=64$ out of 100 realizations of channels, MEB will be able to serve 10 SUs only 27 times, while ZFB can serve them all 100 times. It has been observed that by increasing $M_b$ from 64 to 1024, MEB  can serve all 10 SUs with probability 1 by allocating $P_k = -20$dB per user.

\subsection{Comparison of equal power with LF MEB and LF ZFB}
\begin{figure}
	\centering
	\includegraphics[width= \columnwidth]{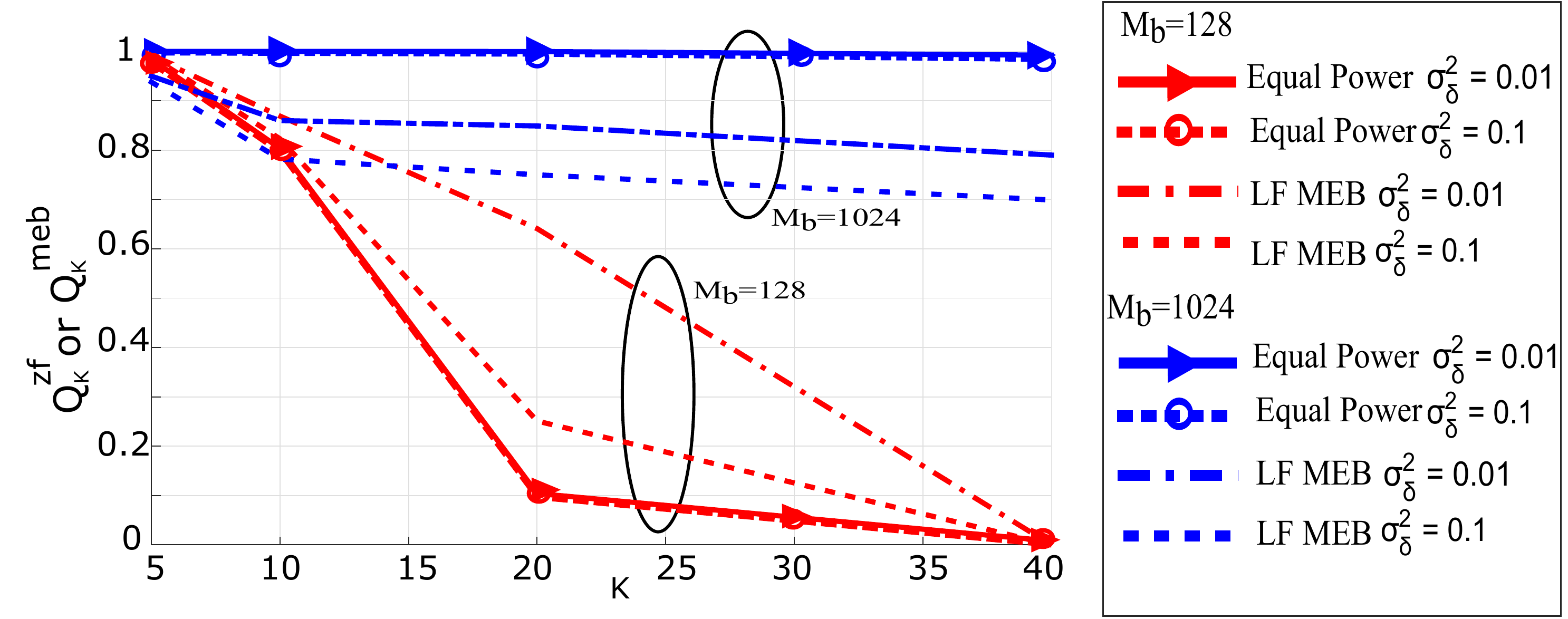}
	\caption{Comparison between equal power MEB and LF MEB. $R^0=1$ bps/Hz, $I^0=-3$dB, $P^0=10$dB.}
	\label{fig:meb_eq_pw_R0_Mb}
	\vspace{-5mm}
\end{figure}

\begin{figure}
	\centering
	\includegraphics[width=  \columnwidth]{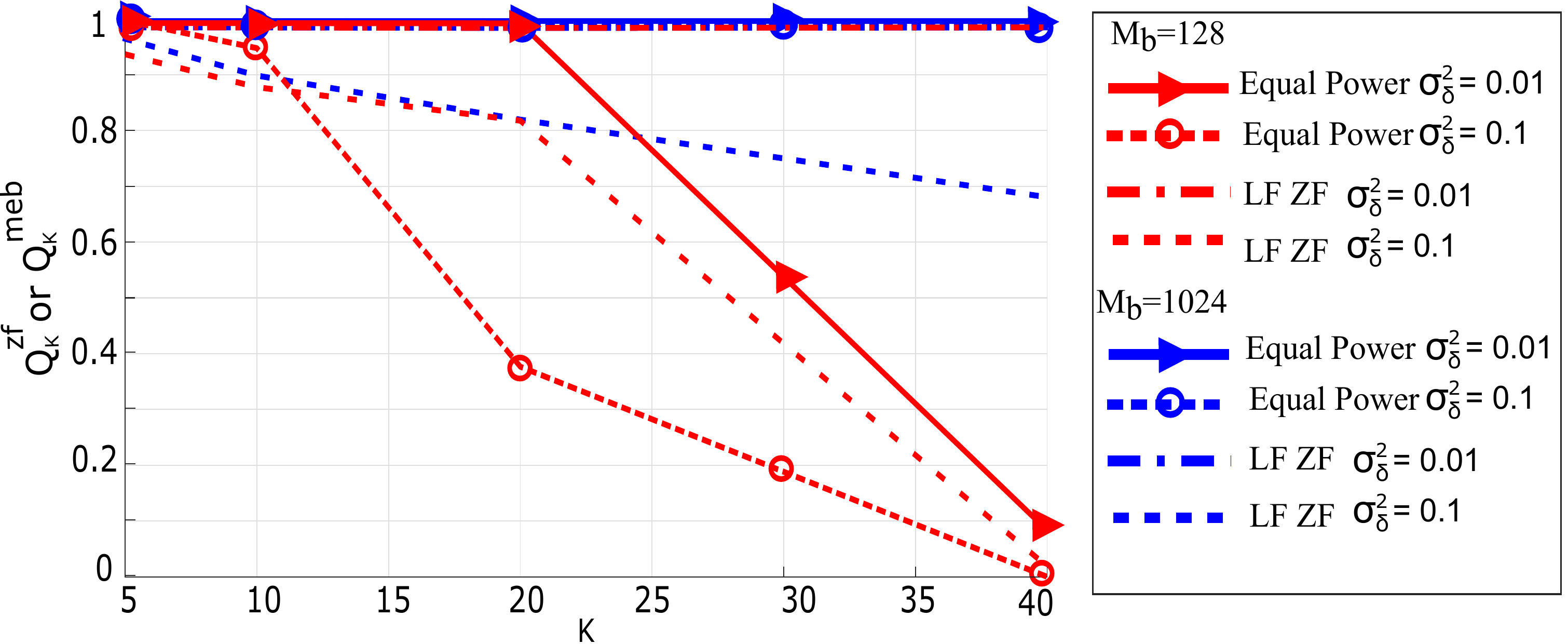}
	\caption{Comparison between equal power ZFB and LF ZFB. $R^0=1$ bps/Hz, $I^0=-3$dB, $P^0=10$dB.}
	\label{fig:zf_eq_pw_R0_Mb}
	\vspace{-5mm}
\end{figure}

The performance of equal power schemes is compared with linear feasibility based power allocation scheme as shown in Fig. \ref{fig:meb_eq_pw_R0_Mb} and \ref{fig:zf_eq_pw_R0_Mb}. For equal power case, we obtain optimal equal power allocation for ZFB and MEB by maximizing $Q^{zf}_{k}$ and $Q^{meb}_{k}$. Therefore, $P^{meb}_{eq}$ and $P^{zf}_{eq}$ are computed before each channel realization, since the equal power allocation depends on channel statistic, $\sigma^2_h$ and $E[\sigma^2_{k,1}]$, and estimation error statistic, $\sigma^2_\delta$. On the other hand, the power allocation using LF ZFB and LF MEB takes into account the estimated channels in each realization. We can observe that the equal power scheme under MEB outperforms LF MEB for $M_b=1024$, while the LF MEB performs better if $M_b=128$ as seen in Fig. \ref{fig:meb_eq_pw_R0_Mb}. Similar trend is observed for ZFB in Fig. \ref{fig:zf_eq_pw_R0_Mb}. Therefore, using LF MEB and LF ZFB power allocation is beneficial if number of antennas is relatively small, $M_b=128$. With large number of antennas, the  precomputed equal power allocation provides higher probability of serving $K$ SUs. The reason for such a behavior is that linear feasibility schemes compute power allocations considering the expected values of SINR and interference, which in turn are functions of variance of channel estimation error (\ref{eq:est_int_to_pu}). The variance does not change with increased number of antennas and therefore overestimates the interference to PUs, which results in reduced probability of serving SUs. The equal power schemes that take into account distributions of interference perform better than linear feasibility schemes.

\begin{figure}[t!]
	\centering
	\begin{subfigure}[b]{0.5\textwidth}
		\centering
		\includegraphics[width = \columnwidth]{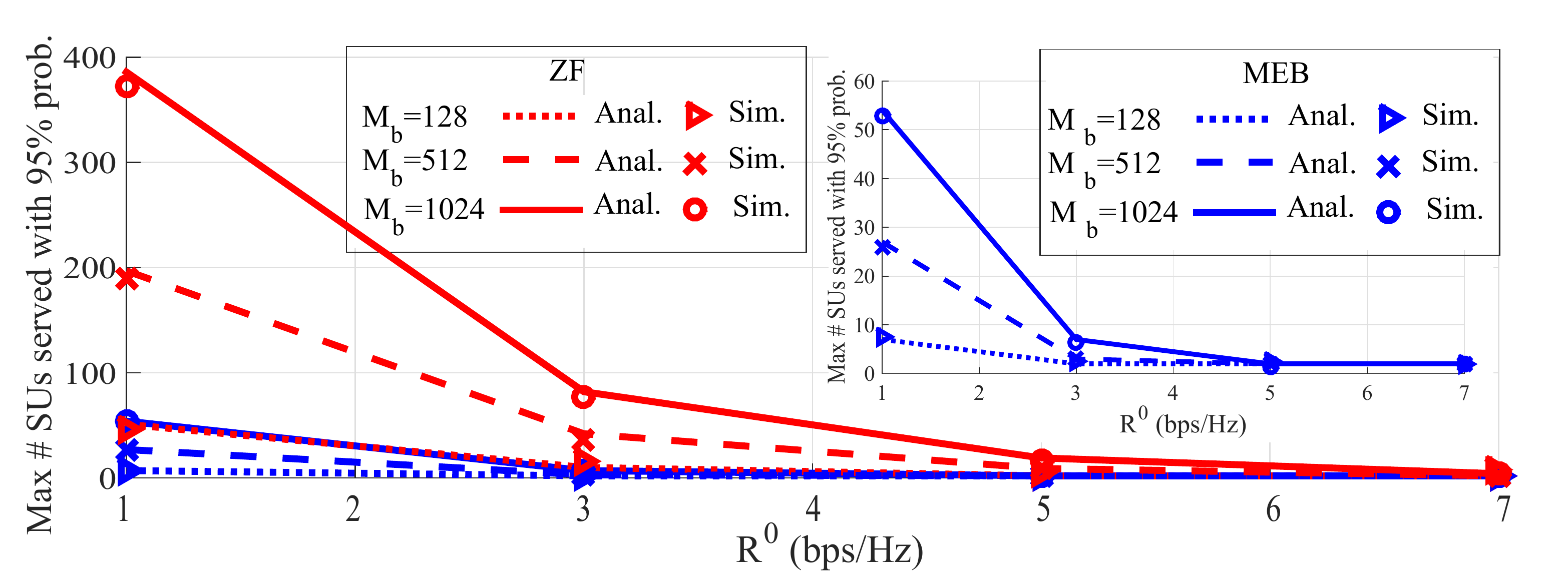}
		\caption{Impact of $R^0$. $I^0=-3$dB, $P^0=10$dB, $\sigma^2_\delta=0.1$}
		\label{fig:R0}
	\end{subfigure}%
	\\
	\begin{subfigure}[b]{0.5\textwidth}
		\centering
		\includegraphics[width = \columnwidth]{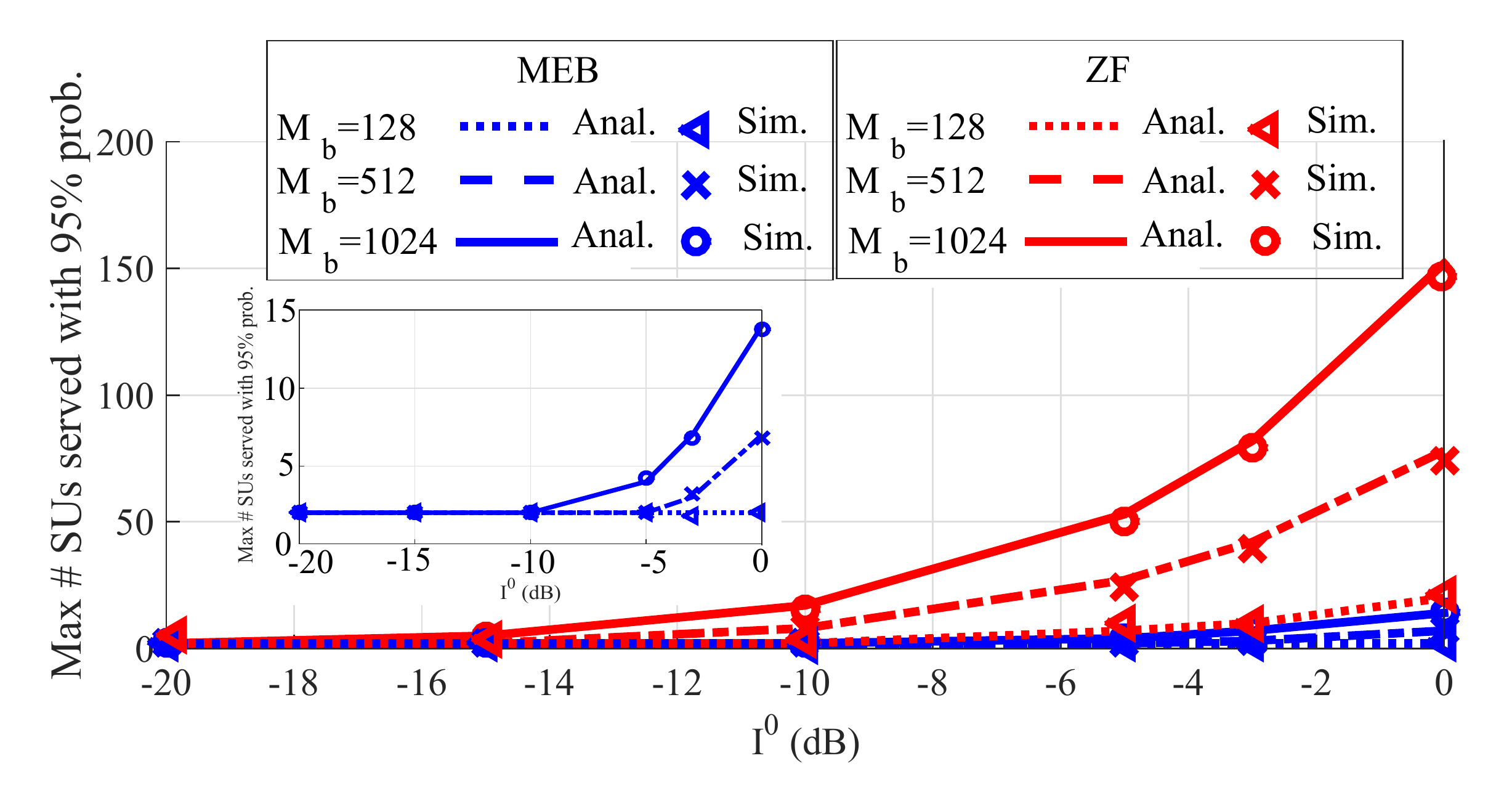}
		\caption{Impact of $I^0$. $R^0=3$ bps/Hz, $P^0=10$dB, $\sigma^2_\delta=0.1$}
		\label{fig:I0}
	\end{subfigure}
	\\
	\begin{subfigure}[b]{0.5\textwidth}
		\centering
		\includegraphics[width = \columnwidth]{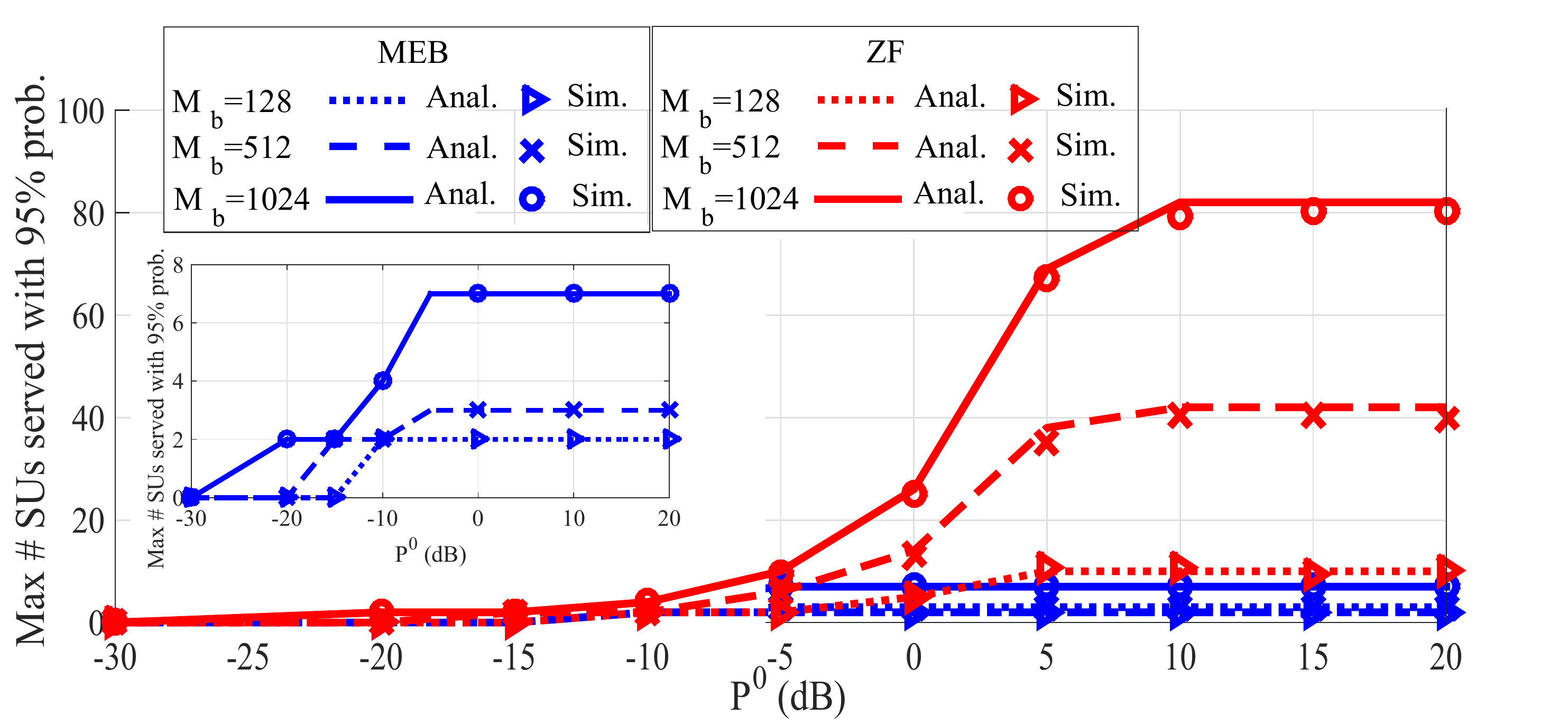}
		\caption{Impact of $P^0$. $I^0=-3$dB, $R^0=3$ bps/Hz, $\sigma^2_\delta=0.1$}
		\label{fig:P0}
	\end{subfigure}		
	\caption{Maximum number of SUs served with 95\% probability.}
	\vspace{-5mm}
	\label{fig:max_sus}
\end{figure}

\subsection{Impact of constraints on equal power scheme}
The impacts of rate, interference and power constraints on MEB and ZFB are shown in Fig. \ref{fig:max_sus}, where the maximum number of SUs served with 95\% probability is plotted against the constraints. We can observe that ZFB always outperforms MEB and serves at most 10x more SUs by canceling out the interference to PUs as well as inter-stream interference. The performance of ZFB and MEB becomes similar as the constraints get more stringent. Although MEB does not perform as well as ZFB, it also does not require the knowledge of CSI between PU and SU. Therefore, in the absence of CSI, the MEB technique can be utilized with equal power allocation. It has been observed, in Fig. \ref{fig:P0}, that the performance of ZFB algorithm saturates if available power is increased beyond 10 dB with 1024 or 512 antennas at SBS. The higher power does not serve more SUs because the gain in SINR is compensated by increased interference to PUs beyond the saturation point.

\section{Conclusions and future work}
\label{sec:Conclusion}
We analyzed the feasibility of serving $K$ SUs in an underlay CR network using massive MIMO base station. We proved that the SINR at SUs can be modeled as a inverse gamma variable under maximum eigenmode beamforming and generalized F distribution under zero forcing beamforming, while the interference to PUs is modeled as a gamma variable under both beamforming schemes. Based on the distributions, the probability of serving $K$ SUs is computed with equal power allocation as a function of given rate, interference and power constraints in the network.

The performance of the equal power allocation is compared with power allocation by solving linear programming feasibility problems. It has been
observed that the linear feasibility based power allocations do
not provide any benefit over equal power scheme if number of
antennas at SBS becomes large. Finally, we observed that ZFB outperforms MEB in terms of serving more SUs under equal power allocation by nullifying the interference to PUs and inter-stream interference. However, ZFB requires the knowledge of the CSI between SUs and PUs to nullify the interference. If no CSI is  available, MEB with equal power allocation can be used to serve at least 1/10th of the SUs as compared to ZFB with same constraints.


In this paper, a feasibility analysis of equal power scheme is presented under a channel propagation where large scale fading coefficients are equal for all channels. This scheme can be extended to a network with unequal large scale fading coefficients by dividing the powers allocated to SUs by corresponding large scale fading coefficients between SBS and SUs. The theoretical framework presented in this paper will be extended in our future work to a more general channel propagation with unequal large scale fading among the channels. We also plan to design SU selection strategies in order to serve maximum number of SUs with given constraints in an underlay CR network with $K$ SUs.

\bibliographystyle{IEEEtran}
\bibliography{IEEEabrv,references}

\appendices

\section{Proof of Theorem 1}
\label{app:thm1}
 Using $\mathbf{v_k} = \mathbf{v_{k,1}}, \mathbf{u_k} = \mathbf{u_{k,1}}$ and $P_k = P^{meb}_{eq}$ in (\ref{eq:sinr_k}) $SINR_k$ for MEB can be expressed as follows:
$
\nonumber SINR^{meb}_k = \frac{1}{\frac{\sigma^2_w}{ P^{meb}_{eq} \sigma^2_{k,1}} + \frac{\sum \limits_{l\in \mathcal{L}_{tx}}P_p |\mathbf{u^H_{k,1}} \mathbf{h_{lk}}|^2}{P^{meb}_{eq} \sigma^2_{k,1}} + \sum \limits_{\mathop{j=1} \limits_{j \neq k}}^{K}|\mathbf{v^H_{k,1}}\mathbf{v_{j,1}}|^2}
= \frac{1}{C + X + Y}
\label{eq:sinr_k_meb}
$

For the simplicity of analysis, we replace $\sigma^2_{k,1}$ with its expected value $E[\sigma^2_{k,1}]$ which is computed using distribution of $\sigma^2_{k,1}$ \cite{Kang2003}.  Therefore, $C$ becomes a constant. It should be noted that each term $|\mathbf{u^H_{k,1}}\mathbf{h_{lk}}|^2$ is the power of the projection of a single beamforming vector $\mathbf{u_{k,1}}$ on an isotropic channel vector $\mathbf{h_{lk}}$. Therefore, $|\mathbf{u^H_{k,1}}\mathbf{h_{lk}}|^2$ is gamma random variable with shape and scale parameters 1 and $\sigma^2_h$, respectively. i.e.  $|\mathbf{u^H_{k,1}}\mathbf{h_{lk}}|^2 \sim \Gamma (1,\sigma^2_h)$ \cite{Hosseini2014}. Using lemma 3 in \cite{Hosseini2014}, $X$ is modeled as a Gamma r.v. with parameters $k_x={L}_{tx}$ and $\theta_x = \frac{P_p \sigma^2_h}{ P^{meb}_{eq} E[\sigma^2_{k,1}]}$. Therefore, $X \sim \Gamma(k_x, \theta_x)$.

Further, the principal right singular vector $\mathbf{v_{j,1}}$  is distributed as $CN(0, \frac{1}{M_b} \textbf{I})$. Therefore, $\mathbf{v_{j,1}}$ is an isotropic vector and $|\mathbf{v^H_{k,1}}\mathbf{v_{j,1}}|^2$ is modeled as Gamma random variable:$|\mathbf{v^H_{k,1}}\mathbf{v_{j,1}}|^2\sim \Gamma(1, \frac{1}{M_b})$. Further, $Y$ is summation of $K-1$ Gamma random variables and can be modeled as Gamma random variable with shape and scale parameters $k_y=K-1$ and $\theta_y = 1/M_b$: $Y \sim \Gamma(k_y, \theta_y)$. Letting $Z = X+Y$, we can model $Z$ as another Gamma random variable: $Z \sim \Gamma(k_z, \theta_z)$. The parameters $k_z$ and $\theta_z$ are computed from $k_x, \theta_x, k_y, \theta_y$ using \textit{lemma 3} in \cite{Hosseini2014}:
{\small
\begin{align}
\nonumber k_z = \frac{\left(\frac{{L}_{tx} P_p \sigma^2_h}{P^{meb}_{eq} E[\sigma^2_{k,1}]} +\frac{K-1}{M_b} \right)^2}{\frac{{L}_{tx} P^2_p \sigma^4_h}{(P^{meb}_{eq})^2 E[\sigma^2_{k,1}]^2} +\frac{K-1}{M^2_b} }~~,
\theta_z = \frac{\frac{{L}_{tx} P^2_p \sigma^4_h}{(P^{meb}_{eq})^2 E[\sigma^2_{k,1}]^2} +\frac{K-1}{M^2_b}}{\frac{{L}_{tx} P_p \sigma^2_h}{P^{meb}_{eq} E[\sigma^2_{k,1}]} +\frac{K-1}{M_b} }
\end{align}
}
Therefore, $SINR^{meb}_{k} = \frac{1}{C+Z}$. The sum $C+Z$ can be modeled as a Gamma random variable with parameters shape and scale parameters $k'$ and $\theta'$, respectively. Using the formula for mean and variance of a Gamma random variable, we get:
{\small
\begin{align}
\small \nonumber E[C+Z] =C + k_z\theta_z = k' \theta', ~~
 var(C+Z) = k_z \theta^2_z = k' (\theta')^2.
\end{align}
}
Solving for $k'$ and $\theta'$ gives $ k' = \frac{(C+k_z \theta_z)^2}{k_z \theta^2_z}$ and $\theta' = \frac{k_z \theta_z^2}{C+k_z \theta_z}$. The parameters are written in terms of $A$ and $B$ as stated in Theorem 3. Therefore, $SINR^{meb}_k$ is reciprocal of a Gamma random variable $C+Z$ and is modeled as a inverse Gamma random variable with shape and scale parameters $k'$ and $1/\theta'$, respectively: $SINR^{meb}_k \sim IG (k',1/\theta')$ and the CDF is given in (\ref{eq:inverse_gamma}).
\section{Proof of Theorem 3}
\label{app:thm3}
In ZFB, SINR expression becomes:
$SINR^{zf}_{k} =\frac{ P^{zf}_{eq} \sigma^2_{k,1} |\mathbf{v^H_{k,1}}\mathbf{v_k}|^2}{\sigma^2_w + \sum_{l\in\mathcal{L}_{tx}}P_p|\mathbf{u^H_{k,1}}\mathbf{h_{lk}}|^2 } = \frac{W}{\sigma^2_w +Q}$. 
The term $|\mathbf{v^H_{k,1}}\mathbf{v_k}|^2$ is power of the projection of $M_b$ dimensional vector $\mathbf{v_{k,1}}$ onto a $M_b-K-L_{tx}+1$ dimensional beamforming space. Note that $\mathbf{v_{k,1}}$ is an isotropic vector with distribution $CN(0, 1/M_b)$. Therefore, according to lemma 1  \cite{Hosseini2014}: $|\mathbf{\mathbf{v^H_{k,1}}}\mathbf{v_k}|^2 \sim \Gamma(M_b-K-L_{tx}+1,1/M_b)$. Approximating $\sigma^2_{k,1}$ with its expected value, we get $W \sim \Gamma(k_n, \theta_n )$, where $k_n = M_b-K-L_{tx}+1$, $\theta_n = \frac{P^{zf}_{eq} E[\sigma^2_{k,1}]}{M_b}$.

In the denominator $Q$ is modeled as Gamma random variable (similar to $X$ in Appendix A), so that $Q \sim \Gamma({L}_{tx}, P_p \sigma^2_h)$. Similar to $C+Z$ in Appendix A, The denominator $\sigma^2_w + Q$ is modeled as Gamma random variable with parameters $k_d = \frac{(\sigma^2_w + {L}_{tx}P_p \sigma^2_h)^2}{{L}_{tx} P^2_p\sigma^4_h}$ and $\theta_d = \frac{{L}_{tx} P^2_p\sigma^4_h}{\sigma^2_w + {L}_{tx}P_p \sigma^2_h}$. Therefore, $SINR^{zf}_k$ is a ratio of two independent Gamma random variables and is modeled as a generalized F distribution \cite{Gia2011} with parameters $k_n,k_d$, and $\lambda = \theta_d/\theta_n = \frac{{L}_{tx} P_p\sigma^2_h M_b}{P^{zf}_{eq}\sigma^2_{k,1}(\sigma^2_w + {L}_{tx}P_p \sigma^2_h)} $. The variable is denoted as $SINR^{zf}_k \sim GF(k_n,k_d,\lambda)$.

\end{document}